\documentclass[prb,twocolumn,floatfix,notitlepage,superscriptaddress,longtable]{revtex4-2}

\usepackage{amsmath, amssymb, amsfonts, bm, esint}
\usepackage{graphicx, epsfig}
\usepackage{psfrag}
\usepackage{overpic}
\usepackage{color}
\usepackage[normalem]{ulem}
\usepackage{lipsum}
\usepackage{verbatim}
\usepackage[figuresright]{rotating}
\usepackage[justification=raggedright,singlelinecheck=false]{caption}
\usepackage[version=4]{mhchem}
\usepackage[svgnames]{xcolor}
\usepackage[colorlinks,linkcolor=blue,anchorcolor=blue,citecolor=blue,urlcolor=blue]{hyperref}

\usepackage{subfig} % <-- kept for RevTeX compatibility (subcaption conflicts)

\usepackage{tikz}
\usetikzlibrary{positioning, arrows.meta, shapes.geometric, calc, backgrounds, fit}

\def\be{\begin{equation}} \def\ee{\end{equation}}
\def\bea{\begin{eqnarray}} \def\eea{\end{eqnarray}}

\def\k{{\bf k}}

\newcommand{\ket}[1]{| #1 \rangle}

\newcommand{\braket}[2]{\langle #1 |#2\rangle}

\definecolor{Qicolor}{RGB}{3, 136, 252}

\usepackage{amsmath}
\usepackage{tikz}
\usetikzlibrary{arrows.meta,calc}

\definecolor{navy}{RGB}{23,42,99}
\definecolor{cellblue}{RGB}{40,82,150}

\newcommand{\cellX}[2]{%
  \draw[navy,line width=0.6pt,fill=cellblue] (#1-0.16,#2-0.16) rectangle (#1+0.16,#2+0.16);}
\newcommand{\cellE}[2]{%
  \draw[navy,line width=0.6pt,fill=white] (#1-0.16,#2-0.16) rectangle (#1+0.16,#2+0.16);}

\definecolor{darkblue}{RGB}{0, 51, 102}
\definecolor{steelcyan}{RGB}{70, 130, 180} 
\definecolor{softgray}{RGB}{240, 240, 245}

\newcommand{\premiumfan}[4]{
    \begin{scope}[shift={(#1,#2)}, scale=#4, transform shape]
        
        \fill[softgray] (0,0) -- (150:1.0) arc (150:30:1.0) -- cycle;
        
        \draw[thin, white, line width=1.5pt] (150:0.75) arc (150:30:0.75);
        \draw[thin, gray!40] (150:0.75) arc (150:30:0.75);

        \pgfmathparse{#3>0.001?1:0}
        \ifnum\pgfmathresult=1
            \pgfmathsetmacro{\endangle}{90 - 60*#3}
            \fill[darkblue!85] (0,0) -- (90:0.95) arc (90:\endangle:0.95) -- cycle;
        \fi
        
        \pgfmathparse{#3<-0.001?1:0}
        \ifnum\pgfmathresult=1
            \pgfmathsetmacro{\absval}{-1*#3}
            \pgfmathsetmacro{\endangle}{90 + 60*\absval}
            \fill[steelcyan!85] (0,0) -- (90:0.95) arc (90:\endangle:0.95) -- cycle;
        \fi

        \draw[thick, darkblue, join=round] (0,0) -- (150:1.0) arc (150:30:1.0) -- cycle;
        
        \foreach \a in {30, 45, 60, 75, 90, 105, 120, 135, 150} {
            \draw[gray!70, thin] (\a:0.95) -- (\a:1.0);
        }
        \foreach \a in {30, 60, 90, 120, 150} {
            \draw[darkblue, thick] (\a:0.9) -- (\a:1.0);
        }
        \draw[thin, darkblue!40] (0,0) -- (90:0.9);

        \pgfmathsetmacro{\needleangle}{90 - 60*#3}
        \begin{scope}[rotate=\needleangle]
            \fill[darkblue] (0, -0.045) -- (0, 0.045) -- (1.2, 0) -- cycle;
        \end{scope}

        \fill[white] (0,0) circle (0.12);
        \draw[thick, darkblue] (0,0) circle (0.12);
        \fill[darkblue] (0,0) circle (0.04);

    \end{scope}
}

\begin{document}
\title{Mechanistic Interpretability and Causal Feature Steering of Neural Quantum States via Sparse Autoencoders}
    
\author{Zihao Qi}
\email[Contact author: ]{zq73@cornell.edu}
\affiliation{Department of Physics, Cornell University, Ithaca, NY 14853, USA.}

\author{Christopher Earls}
\affiliation{Center for Applied Mathematics, Cornell University, Ithaca, NY 14853, USA.}

\date{\today}

\begin{abstract}
Neural Quantum States (NQS) are a remarkably expressive class of variational ansätze for quantum many-body wavefunctions, yet little is understood about their internal mechanisms: trained on variational objectives alone, how do NQS accurately capture physical observables that they have never been explicitly optimized for? In this work, we present a systematic approach to analyze the internal activations of NQS using sparse autoencoders. We extract features from the residual stream and demonstrate that these features strongly correlate with physical observables such as order parameters, staggered magnetization, and half-chain correlators, across both ground state representation and real-time dynamics. Remarkably, the discovery of these features is entirely unsupervised, with no physical labels provided. We further establish that such features causally affect the corresponding observables predicted by NQS, by showing that targeted, post-training intervention on a \textit{single} feature smoothly and monotonically steers the corresponding observable, while leaving the variational energy nearly unchanged. These results demonstrate that NQS are not merely functional approximators, but encode rich, interpretable internal representations of physical information. Our approach provides both a diagnostic and an intervention tool for NQS, and serves as a foundation for using mechanistic interpretability towards more reliable, transparent NQS.
\end{abstract}

\maketitle

\section{Introduction}
Neural quantum states (NQS) have recently emerged as a powerful class of variational ansätze for quantum many-body systems~\cite{lange_review, nqs_review, nqs_review2, carleo2017}. By leveraging the expressivity of neural networks to parameterize many-body wavefunctions, NQS have achieved remarkable success across spin and fermionic systems, in both ground-state representation and real-time dynamics~\cite{carleo2017,real_time_dynamics1,real_time_dynamics2,time_dependent_nqs1, time_dependent_nqs2, RNN_NQS, transformer_frustrated, transformer_nqs_zhang, nqs_magic, nqs1, nqs_2026, NQS2, NQS3, autoregressive_NQS, autoregressive_nqs2, thermodynamic_limit, nqs_fuliang2, NQS_Fuliang}.

The expressivity of NQS, however, comes at the expense of transparency. Despite their empirical success, NQS remain largely ``black boxes''. This raises a fundamental question: trained on variational objectives such as energy minimization alone, how do NQS accurately capture physical observables that have never been explicitly provided as optimization targets? Is such physical information somehow represented inside the network, and if so, where and in what form? Can such representations be extracted, interpreted, or causally manipulated? At present, such questions remain largely unanswered.

This opacity is not only conceptual, but also directly impacts practical aspects of NQS, especially regarding model trust and robustness. In many-body quantum systems, NQS are often used precisely in regimes where obtaining numerical benchmarks can be challenging. In such settings, one would also like to have independent, physically motivated probes on NQS models to assess whether the network has learned a physically sensible representation of the quantum state and whether its predictions can be validated and trusted.

Existing works have mostly approached the opacity of NQS \textit{externally}, by imposing physically motivated architectural constraints~\cite{NQS_design_choice, NQS_symmetry2_Luodi, NQS_Symmetry, NQS_Symmetry3_Luodi, towards_nqs_interp, nqs_interp_correlation, nqs_lack_interp_transformer, autoregressive_nqs_enforce_symmetry, nqs_2026, thermodynamic_limit, NQS_symmetry_1}, understanding their capacity to represent quantum states~\cite{NN_representability, NQS_entanglement_bound_nisarga, NQS_volume_law, nqs_volume_law_comment, nqs_representability_2017, nqs_representability_2018}, and analyzing model parameters~\cite{scalinglaw, model_param1, model_param2, model_param3, double_descent, model_param4}. These efforts ask what physical structure should be built into, or can be represented by, neural quantum states, and are essential for designing more physically faithful NQS. However, a \textit{post hoc} analysis of trained NQS models, which would reveal how networks \textit{internally} organize physical information, remains missing. Probing the internal representations formed by NQS is important not only for understanding how neural networks represent quantum states, but also for building more informed and reliable variational tools for quantum many-body problems.

In this work, we present a systematic framework to bridge this gap using tools from mechanistic interpretability, a rapidly developing field that seeks to interpret neural networks by analyzing their activations and response to causal interventions~\cite{mech_interp_review1, mech_interp_review2, mech_interp_review3}. In particular, we use a sparse autoencoder (SAE) to decompose NQS activations into a sparse set of interpretable features. Originally applied to large language models~\cite{LLM_SAE1, LLM_SAE2, LLM_SAE3, claude_interpretable, sparse_coding}, SAEs have also successfully extracted various scientific concepts from protein language models and weather predictors~\cite{ProteinLM_SAE1, ProteinLM_SAE2, PLM_interp4, PLM_steering_interp3, weather_model, SAE_fluiddynamics}. 

Compared to these models, NQS offer a particularly ideal setting for interpretability analysis. NQS are not trained to learn patterns from an external dataset, but rather by minimizing physical objectives such as the variational energy. As a result, any structure found inside the network reflects how NQS internally organize information in order to solve quantum many-body problems such as ground state search, as opposed to being inherited from an external dataset. Furthermore, unlike feature labels in many language and scientific models, which are often semantic and difficult to validate uniquely, the relevant concepts in NQS are precisely defined physical observables. This allows feature interpretations to be tested quantitatively, through both observable correlations and causal interventions.

To benchmark our approach, we first analyze the internal representations of transformer-based NQS for both ground state of, and time evolution under, the 1D Transverse-Field Ising Model (TFIM). Applying SAEs to the final-layer residual stream, we extract sparse features that correlate nearly perfectly with physical observables, including the order parameter, absolute magnetization, and half-chain spin correlator. Remarkably, such feature learning is entirely autonomous: the SAE is not, \textit{a priori}, informed of what features to look for; as a result, the discovery indicates that the model has formed its own internal representations of such physical observables during training. We further establish that the features play a \textit{causal} role in predicting the physical observables through targeted intervention: rescaling the sparse feature smoothly steers the corresponding observable, while leaving the variational energy almost unchanged. We extend this analysis to the two-dimensional Heisenberg antiferromagnetic model, demonstrating that the approach generalizes beyond one-dimensional and exactly solvable systems.

Our results demonstrate that NQS do not merely approximate wavefunction amplitudes configuration by configuration, but rather organize physical quantities into their internal coordinates. This provides both a diagnostic tool for assessing whether trained NQS models have learned physically meaningful representations, as well as a route toward controlled, feature-level manipulation of NQS. More broadly, our approach opens the door to a two-way exchange between quantum many-body systems and mechanistic interpretability: NQS provide a setting in which interpretability claims and tools can be tested against ground truths; conversely, interpretability offers methods and probes to understand, diagnose, and ultimately improve the reliability of NQS models.

The rest of this work is organized as follows. In Sec.~\ref{sec:background}, we review the basics of neural quantum states (NQS) and sparse autoencoders (SAE), a tool from the interpretability community that extracts patterns from neural network activations. In Sec.~\ref{sec:results}, we demonstrate that SAE is able to extract sparse features in NQS residual streams that strongly correlate with, and causally control, physical observables, for both ground state and time-evolved states. We extend our analysis to the 2D Heisenberg Antiferromagnetic model in Sec.~\ref{sec:heisenberg_afm}, demonstrating the method's applicability beyond simple systems.
We summarize our results and discuss directions for future work in Sec.~\ref{sec:discussions}.

\section{Background~\label{sec:background}}
\subsection{Neural Quantum States \label{sec:NQS}}
In this section, we briefly review neural quantum states (NQS), a neural-network-based variational ansätz for quantum many-body wavefunctions. For concreteness, we focus on systems with spin-$1/2$ degrees of freedom.
The computational basis consists of tensor products of
Pauli-$z$ eigenstates,
$\ket{\boldsymbol{\sigma}}
 = \ket{\sigma_1,\sigma_2,\ldots,\sigma_L}$,
where $L$ is the system size and $\sigma_i\in\{-1,+1\}$ denotes the eigenvalue of the Pauli-$z$ operator on site $i$.

A quantum many-body wavefunction, therefore, can be viewed as a mapping from the configuration space (here bit-strings of length $L$: $\{-1, 1 \}^L$) to the associated complex amplitude:
\begin{equation}
    \psi: \boldsymbol \sigma \rightarrow \braket{\boldsymbol \sigma}{\psi} \in \mathbb{C}.
\end{equation}

The central idea of NQS is to leverage the expressivity of neural networks to parametrize the function $\psi$, while using only a polynomial number of parameters $\boldsymbol \theta$. Among the neural network architectures proposed for this purpose, we focus on autoregressive quantum states~\cite{RNN_NQS, transformer_frustrated, transformer_nqs_zhang, autoregressive_NQS, autoregressive_nqs2}, and in particular transformer-based representations. The transformer architecture used in this work is illustrated in Fig.~\ref{fig:illustration}(a), with implementation details in Appendix~\ref{app:transformer_architecture}.

In practice, the autoregressive ansätze represent the wavefunction amplitude $\psi(\boldsymbol{\sigma})$ as a Born probability $p(\boldsymbol{\sigma})$ and a phase $\phi(\boldsymbol{\sigma})$, which are output separately:
\begin{equation}
  \psi_{\boldsymbol{\theta}}(\boldsymbol{\sigma})
  = \sqrt{p_{\boldsymbol{\theta}}(\boldsymbol{\sigma})}\;
    e^{i\phi_{\boldsymbol{\theta}}(\boldsymbol{\sigma})}.
  \label{eq:mod_phase}
\end{equation}

In the autoregressive transformer wavefunction ansätze, the Born probability distribution is factorized into a product of conditional probabilities,
\begin{equation}
  p_{\boldsymbol{\theta}}(\boldsymbol{\sigma})
  = \prod_{i=1}^{L}
    p_{\boldsymbol{\theta}}
    (\sigma_i\,|\,\sigma_1,\ldots,\sigma_{i-1}),
  \label{eq:factorization}
\end{equation}
where $p(\sigma_i\,|\,\sigma_1,\ldots,\sigma_{i-1})$ is the conditional probability of the $i$th spin, given the configuration of all preceding spins. The transformer therefore acts as a causal sequence model for spin configurations: at each site, the distribution of the next spin is predicted, conditioning on all previously generated spins. The phase is also decomposed analogously.

Since each conditional probability is normalized, the autoregressive factorization ensures that the full Born distribution is normalized by construction. This factorization also enables
\emph{exact} autoregressive sampling: each configuration is generated sequentially site-by-site, via sampling from the corresponding conditional probabilities. All configurations are independently drawn from the Born distribution $|\psi_{\boldsymbol{\theta}}|^2$, avoiding Markov-chain inter-sample autocorrelation.

From such samples, expectation values of an observable $O$ can be estimated by Monte Carlo,
\begin{equation}
  \langle O\rangle
  = \mathop{\mathbb{E}}_{\boldsymbol{\sigma}\sim
      |\psi_{\boldsymbol{\theta}}|^2}
    \bigl[O_{\mathrm{loc}}(\boldsymbol{\sigma})\bigr],
\end{equation}    
where the local estimator
\begin{equation}
  O_{\mathrm{loc}}(\boldsymbol{\sigma})
  = \sum_{\boldsymbol{\sigma}'}
    \braket{\boldsymbol{\sigma}}{O |\boldsymbol{\sigma'}}
    \frac{\psi_{\boldsymbol{\theta}}(\boldsymbol{\sigma}')}
         {\psi_{\boldsymbol{\theta}}(\boldsymbol{\sigma})}
  \label{eq:local_estimator}
\end{equation}
can be efficiently evaluated when $O$ is sparse in the computational basis. 

In particular, to represent the ground state, the NQS parameters are optimized to minimize the variational energy
\begin{equation}
  E(\boldsymbol{\theta})
  = \frac{\langle \psi_{\boldsymbol{\theta}}|H |
{\psi_{\boldsymbol{\theta}}} \rangle } {\braket{\psi_{\boldsymbol{\theta}}}    {\psi_{\boldsymbol{\theta}}}}.
  \label{eq:var_energy}
\end{equation}
In practice, both the energy and its gradient are estimated stochastically during training.

NQS can also be used to simulate real-time dynamics through time-dependent parameters $\boldsymbol \theta(t)$. The goal is to approximate exact Schr\"odinger evolution within the variational manifold. The network parameters are optimized so that the variational residual
\begin{equation}
  R(\dot{\boldsymbol{\theta}})
  = \Bigl\|\,
      \sum_k \dot{\theta}_k\,
      \partial_{\theta_k}\ket{\psi_{\boldsymbol{\theta}}}
      + iH\ket{\psi_{\boldsymbol{\theta}}}
    \,\Bigr\|
  \label{eq:residual}
\end{equation}
is minimized at each time step.
This yields the time-dependent
variational principle (TDVP) equation of motion,
\begin{equation}
  \sum_{k'} S_{kk'}\,\dot{\theta}_{k'} = -i\,F_k,
  \label{eq:tdvp}
\end{equation}
where
$S_{kk'} = \langle O_k^{*}O_{k'}\rangle
          -\langle O_k^{*}\rangle\langle O_{k'}\rangle$ is the covariance matrix,
and
$F_k = \langle O_k^{*}E_{\mathrm{loc}}\rangle
-\langle O_k^{*}\rangle\langle E_{\mathrm{loc}}\rangle$ is the force vector. The quantity $O_k = \partial_{\theta_k}\log\psi_{\boldsymbol{\theta}}$ is the logarithmic derivative of the wavefunction. Both $S_{k k'}$ and $F_k$ are Monte-Carlo estimates over samples from the instantaneous Born distribution
$|\psi_{\boldsymbol{\theta}(t)}|^2$. In what follows, we apply sparse autoencoders to analyze the internal representations formed by NQS in both settings.

\begin{figure*}[htbp]
\centering
\begin{tikzpicture}[
  font=\small,
  >={Stealth[length=2.4mm,width=2.0mm]},
  arr/.style={->,line width=0.9pt},
  box/.style={draw=navy,line width=0.9pt,fill=white,align=center,inner sep=4pt},
  layer/.style={draw=navy,line width=0.8pt,fill=white,minimum width=1.5cm,minimum height=0.42cm},
]

%================= PANEL (a) : Wave-function ansätze =================%
\def\xa{0}
\node[align=center,font=\large] at (\xa,9.85) {Wavefunction ansätze};
\node[font=\LARGE] at (\xa-2.55,9.85) {\textbf{(a)}};
\node[font=\Large] at (\xa+0.15,9.25) {$(\sigma_1,\sigma_2,\ldots,\sigma_L)$};

\node[box,minimum width=2.6cm,minimum height=0.55cm] (emb) at (\xa,8.25) {Embedding};
\draw[arr] (\xa,9.05) -- (emb.north);

\node[draw=navy,line width=0.9pt,minimum width=3.0cm,minimum height=3.3cm] (nn) at (\xa,5.9) {};
\node[align=center,anchor=east] at (nn.west) {Transformer\\Layers};

\node[layer] (L1) at (\xa,7.00) {Layer 1};
\node[layer] (L2) at (\xa,6.05) {Layer 2};
\node at (\xa,5.42) {$\vdots$};
\node[layer] (L3) at (\xa,4.75) {Layer $L_T$};
\draw[arr] (emb.south) -- (L1.north);
\draw[arr] (L1.south) -- (L2.north);

\node[box,minimum width=2.6cm,minimum height=0.55cm] (unemb) at (\xa,3.15) {Unembedding};
\draw[arr] (L3.south) -- (unemb.north);
\draw[arr] (unemb.south) -- (\xa,2.3);
\node[font=\Large] at (\xa,1.95) {$\psi(\sigma)$};

% tap the residual stream (just before the unembedding layer) -> h, zig-zag
\fill[navy] (\xa,3.97) circle (1.3pt);
\draw[arr] (\xa,3.97) -- (1.95,3.97) -- (1.95,6.0) -- (3.95,6.0);

\node[font=\normalsize] at (2.8,6.25) {Residual};
\node[font=\normalsize] at (2.8,5.75) {Stream};

%================= PANEL (b) : Latent representation =================%
\def\yb{6.0}
\node[align=center,font=\large] at (5.75,9.85) {Sparse Autoencoder};
\node[font=\LARGE] at (3.35,9.85) {\textbf{(b)}};

% input bar h
\draw[navy,line width=0.9pt,fill=white] (3.97,\yb-1.10) rectangle (4.23,\yb+1.10);
\node[font=\Large] at (4.10,\yb-1.65) {$\mathbf{h}^{(m)}$};

% encoder: short on the h side, very tall on the SAE side (opens wide toward z)
\draw[navy,line width=0.9pt,fill=white]
      (4.67,\yb-0.45)--(5.17,\yb-1.70)--(5.17,\yb+1.70)--(4.67,\yb+0.45)--cycle;
\node[font=\normalsize] at (4.95,\yb + 2.00) {Encoder};

% latent vector z : long gapless column, only 3 lit
\def\xz{6.00}
\foreach \k in {-8,...,8}{\cellE{\xz}{\yb+0.32*\k}}
\cellX{\xz}{\yb+0.32*5}
\cellX{\xz}{\yb+0.32*1}
\cellX{\xz}{\yb-0.32*4}
\cellX{\xz}{\yb-0.32*7}
\node[align=center,font=\normalsize] at (\xz,\yb-3.35) {Sparse\\Activations $\mathbf{z}^{(m)}$};
\node[align=center,font=\large] at (\xz,\yb-4.15) {$\mathcal{L} = \sum_m ||\mathbf{h}^{(m)}-\mathbf{h'}^{(m)}||_2^2 +\lambda ||\mathbf{z}^{(m)}||_1$};

% decoder: very tall on the SAE side, short on the h' side (narrows toward h')
\draw[navy,line width=0.9pt,fill=white]
      (6.83,\yb-1.70)--(7.33,\yb-0.45)--(7.33,\yb+0.45)--(6.83,\yb+1.70)--cycle;
\node[font=\normalsize] at (7.08,\yb+2.00) {Decoder};

% output bar h'
\draw[navy,line width=0.9pt,fill=white] (7.90,\yb-1.10) rectangle (8.16,\yb+1.10);
\node[font=\Large] at (8.1,\yb-1.65) {$\mathbf{h'}^{(m)}$};

% arrows of panel (b)
\draw[arr] (4.23,\yb) -- (4.62,\yb);
\draw[arr] (5.22,\yb) -- (\xz-0.22,\yb);
\draw[arr] (\xz+0.22,\yb) -- (6.78,\yb);
\draw[arr] (7.38,\yb) -- (7.86,\yb);

%================= PANEL (c) : Observable readout =================%
\def\xc{10.2}
\node[align=center,font=\large] at (\xc+0.65,9.85) {Feature Tuning $\rightarrow$};
\node[align=center,font=\large] at (\xc+0.65,9.35) {Observable Control};

\node[font=\LARGE] at (\xc-1.8,9.85) {\textbf{(c)}};

% long, gapless column (12 cells), 3 lit (one per readout), vdots + trailing cell
\def\xz{10.40}
\foreach \k in {-8,...,8}{\cellE{\xz}{\yb+0.32*\k}}
\cellX{\xz}{\yb+0.32*5}
\cellX{\xz}{\yb+0.32*1}
\cellX{\xz}{\yb-0.32*4}
\cellX{\xz}{\yb-0.32*7}

% 1. Very Small Slider (Scale = 0.4)

    \premiumfan{9.5}{7.4}{-0.75}{0.5}

    \premiumfan{9.5}{6.1}{+0.3}{0.5}

    \premiumfan{9.5}{4.5}{+0.8}{0.5}

% readout arrows + labels
\draw[arr] (\xc+0.22,\yb-0.32*4) -- (\xc+1.40,\yb-0.32*4);
\node[anchor=west,font=\large] at (\xc+1.45,\yb-0.32*4) {$M_\text{stag}$};
\draw[arr] (\xc+0.22,\yb+0.32*5) -- (\xc+1.40,\yb+0.32*5);
\node[anchor=west,font=\large] at (\xc+1.45,\yb+0.32*5) {$C_{zz}$};
\draw[arr] (\xc+0.22,\yb+0.32*1) -- (\xc+1.40,\yb+0.32*1);
\node[anchor=west,font=\large] at (\xc+1.45,\yb+0.32*1) {$\langle M_z(t)\rangle$};
\draw[arr] (\xc+0.22,\yb-0.32*7) -- (\xc+1.40,\yb-0.32*7);
\node[anchor=west,font=\large] at (\xc+1.45,\yb-0.32*7) {$\cdots$};

\end{tikzpicture}
    \caption{\textbf{Illustration of the workflow of interpreting Neural Quantum States (NQS) using Sparse Autoencoders (SAE).} \textbf{(a)} Transformer-based NQS takes spin configurations $(\sigma_1, \sigma_2, \dots, \sigma_L)$ as input. The tokens are embedded to a latent space, sequentially passed through $L_T$ transformer layers, and unembedded to yield the wavefunction amplitude $\psi(\sigma)$. The final-layer residual stream $\mathbf{h}^{(m)}$ is collected and passed to the Sparse Autoencoder (SAE). \textbf{(b)} The SAE encodes each activation vector $\mathbf{h}^{(m)}$ into a sparse representation $\mathbf{z}^{(m)}$, and reconstructs $\mathbf{h'}^{(m)}$ via a decoder. The SAE is trained to minimize the combined reconstruction and sparsity loss, thereby representing each activation vector as a sparse linear combination of feature directions. \textbf{(c)} The sparse feature directions that SAE identifies not only strongly correlate with observables such as staggered magnetization, half-chain correlators, and time-dependent magnetization, but in fact causally control them. Tuning the feature strengths monotonically changes the corresponding observables.}
    \label{fig:illustration}
\end{figure*}
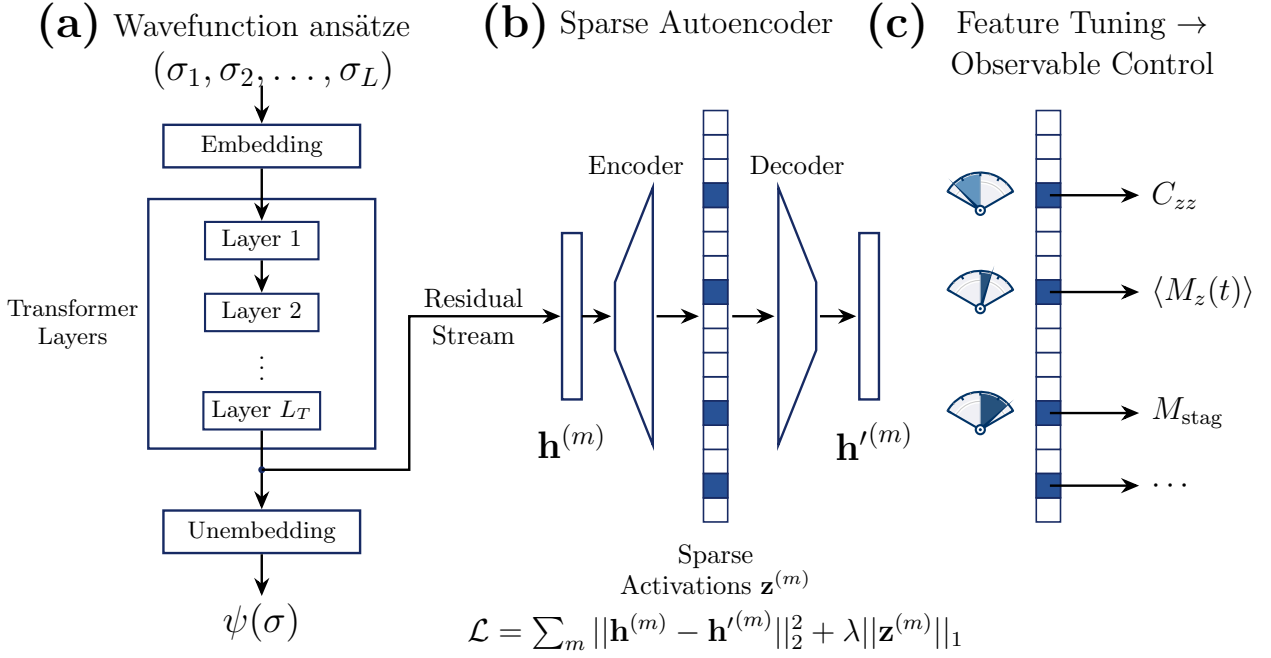

\subsection{Sparse Autoencoders \label{sec:SAE}}
A sparse autoencoder (SAE) embodies a dictionary learning approach~\cite{sparse_coding} that decomposes activation vectors of neural networks into sparse combinations of interpretable learned feature directions. 
%\Chris{we may want to think about the naming conventions that we apply here, in Appendix A, and also incidentally all throughout the manuscript}.
This decomposition is motivated by the \textit{superposition hypothesis}~\cite{superposition_hypothesis}, which postulates that neural networks with residual-stream dimension $d$ can encode far more than $d$ distinct features by superimposing the features in activation space. The SAE approach is designed to disentangle this superposition; it has successfully identified many human-interpretable features in large language models~\cite{LLM_SAE1, LLM_SAE2, LLM_SAE3} and more recently scientific concepts in the natural sciences. For example, SAEs have extracted biological motifs such as catalytic-site activation in protein language models~\cite{ProteinLM_SAE1, ProteinLM_SAE2, PLM_steering_interp3, PLM_interp4}, cyclone-related features in weather models~\cite{weather_model}, and enstrophy-related features in fluid-mechanical foundation models~\cite{SAE_fluiddynamics}.

In this work, the inputs to the SAE are activation vectors within the residual stream of the transformer NQS, $\mathbf{h}^{(m)} \in \mathbb{R}^{d_\text{model}}$. Here $d_\text{model}$ is the dimension of the transformer residual-stream and $m$ indexes the input. The SAE learns a dictionary $\{ \mathbf{f}_k \}$, where $k \in \{1, 2, \dots, d_{\text{SAE}} \}$. Each ``feature" $\mathbf{f}_k \in \mathbb{R}^{d_\text{model}}$ is unit-normalized, and the dictionary is \textit{overcomplete}, in the sense that its number of elements exceeds the residual stream dimension, $d_\text{SAE} > d_\text{model}$. The SAE is trained to find both the feature directions $\mathbf{f}_k$ and sparse, input-dependent coefficients $z_k^{(m)}$, such that each activation vector $\mathbf{h}^{(m)}$ can be approximated as 
\begin{equation}
    \mathbf{h}^{(m)} \simeq \sum_k z_k^{(m)} \, \mathbf{f}_k,
\end{equation}
with only a small number of non-zero coefficients $z_k^{(m)}$. Thus, the SAE represents each activation vector as a sparse linear combination of feature directions.

Architecturally, the SAE is an autoencoder with a single hidden layer of dimension $d_\text{SAE}$. In contrast to conventional autoencoders~\cite{autoencoder}, which generally compress inputs down to lower dimensions, the hidden layer in SAEs is deliberately \emph{higher dimensional} than its input. The SAE first lifts an activation vector $\mathbf{h^{(m)}}\in\mathbb{R}^{{d_\text{model}}}$ to $\mathbf{z}^{(m)}\in\mathbb{R}^{d_{\mathrm{SAE}}}$ via an encoder:
\begin{equation}
\mathbf{z}^{(m)} = \sigma\!\left(W_{\mathrm{enc}} \, \mathbf{h}^{(m)}+\mathbf{b}_{\mathrm{enc}}\right),
\label{eq:SAE_encoding}
\end{equation}
where $W_{\mathrm{enc}}$ is the encoder weight matrix, $\mathbf{b}_{\mathrm{enc}}$ the encoder bias, and $\sigma$ a pointwise non-linear activation function. The SAE then projects $\mathbf{z}^{(m)}$ back to a reconstructed activation ${\mathbf{h'}^{(m)}}\in\mathbb{R}^{d_\text{model}}$ via a decoder:
\begin{equation}
{\mathbf{h'}}^{(m)} = W_{\mathrm{dec}}\,\mathbf{z}^{(m)} + \mathbf{b}_{\mathrm{dec}} = \sum_{k=1}^{d_{\mathrm{SAE}}} z_k^{(m)}\,\mathbf{f}_k + \mathbf{b}_{\mathrm{dec}},
\label{eq:SAE_decoding}
\end{equation}
where the columns of $W_{\mathrm{dec}}$ are normalized feature directions $\mathbf{f}_k$. The encoder weights $W_{\mathrm{enc}}$, both biases $\mathbf{b}_{\mathrm{enc}},\mathbf{b}_{\mathrm{dec}}$, and features (or equivalently $W_{\mathrm{dec}}$) are all learned during training. The concept and architecture of SAEs are illustrated in Fig.~\ref{fig:illustration}(b).

The SAE is trained by minimizing the following loss function summed over all residual-stream activations:
\begin{equation}
  \mathcal{L}_\text{SAE} = \sum_m \left( \|\mathbf{h}^{(m)} - \mathbf{h}'^{(m)}\|_2^2 + \lambda\|\mathbf{z}^{(m)}\|_1 \right),
\end{equation}
where $\| \cdot \|_2$ is the Euclidean ($L_2$) norm, and $\|\cdot \|_1$ is the $L_1$ norm. The first term enforces faithful reconstruction of the activation vectors, while the second term imposes a sparsity penalty that forces all but a few entries of $\mathbf{z}^{(m)}$ to vanish. The competition between the two terms in the loss function ensures that the network finds the right balance between reconstruction fidelity and sparsity. The hyperparameter $\lambda$ controls the relative scale between the two terms. Details about the SAE hyperparameters are in Appendix~\ref{app:nqs_sae_training}.

Importantly, the SAE is trained in an entirely unsupervised manner on the activation vectors, without any access to physical labels. The feature directions must be discovered solely from the transformer's internal representations. As we demonstrate in the next two sections, despite this seemingly blind decomposition, the SAE recovers features that correlate strongly with physical observables.

\section{Results: Transverse Field Ising Model \label{sec:results}}
We start with the one-dimensional Transverse-Field Ising Model (TFIM)~\cite{TFIM}, with the Hamiltonian:
\begin{equation}
  H = -J\sum_{i=1}^{L} \sigma^z_i \sigma^z_{i+1} - h\sum_{i=1}^{L} \sigma^x_i,
\end{equation}
where $\sigma^x_i$ and $\sigma^z_i$ are the Pauli $x$- ($z$-) operators on site $i$, $J$ is the nearest-neighbor Ising coupling, and $h$ the transverse field strength. We impose periodic boundary conditions for a chain of $L=40$ sites. The TFIM has a critical point at $h/J = 1$: for $h < J$, the system is ferromagnetically ordered; with $h > J$, the system is paramagnetic. We set $J \equiv 1$ throughout the rest of this work.

The TFIM provides an ideal starting point for mechanistically interpreting NQS. The model is exactly solvable and well-understood, and therefore provides a controlled setting in which the ground truth is known. For example, the ground state energy density can be exactly solved given $J$ and $h$, which allows us to assess convergence of NQS training. Furthermore, because the relevant physical observables are known, we can directly test whether the features autonomously extracted by SAE genuinely correspond to physically meaningful quantities.

In this section, we study transformer-NQS representations of both the TFIM ground state and real-time dynamics. In the next section, we turn to a two-dimensional, non-integrable model, the 2D Heisenberg Antiferromagnet, to demonstrate that our approach is not restricted to exactly solvable one-dimensional systems.

\subsection{Feature Identification for Ground State \label{sec:gs_correlation}}
We begin by applying the SAE analysis to transformer-based representations of the TFIM ground state. As introduced in Sec.~\ref{sec:NQS}, the transformer is trained to model the ground state by stochastically minimizing the variational energy. Details about the transformer architecture and model parameters are in Appendix~\ref{app:transformer_architecture} and Appendix~\ref{app:nqs_sae_training}, respectively.

After training, we autoregressively sample $M$ spin configurations $\boldsymbol\sigma^{(m)}$ (with $m\in\{1,2,\dots,M\}$) from $|\psi(\boldsymbol\sigma)|^2$. Because the wavefunction is represented autoregressively, these spin samples are drawn from the Born distribution independently, without Monte-Carlo autocorrelation. For each configuration, we store the final-layer residual stream activations of the transformer as a matrix $\mathbf{h}^{(m)}\in\mathbb{R}^{L\times d_\text{model}}$, where $L$ is the system size and $d_\text{model}$ is the residual-stream dimension. The row vector $\mathbf{h}_i^{(m)}$ is the residual stream activation on site $i$ for the $m$th configuration. We focus on the final-layer residual stream, because these activations enter the unembedding layer (Fig.~\ref{fig:illustration}(a)) and directly determine the model's output. 

The SAE is trained on the collection of site-level residual stream activations, using the architecture and loss function as described in Sec.~\ref{sec:SAE}. After training, the SAE decomposes each activation into a sparse linear combination of feature directions
\begin{equation}
    \mathbf{h}_i^{(m)} \approx \sum_k z_{ik}^{(m)} \, \mathbf{f}_k,
\end{equation}
where the coefficient $z_{ik}^{(m)}$ denotes the strength of the $k$th feature, on the $i$th site, for the $m$th configuration.

To associate each SAE feature with a single number per spin configuration, we ``mean-pool'' over the spin chain and average the activation over all sites,
\begin{equation}
  Z_k^{(m)} = \frac{1}{L}\sum_{i=1}^L z_{ik}^{(m)}.
\end{equation}
Thus, $Z_k^{(m)} \in \mathbb{R}$ is a \textit{scalar} feature strength. Collecting $Z_k^{(m)}$ across all configurations yields a vector $\mathbf{Z}_k \in \mathbb{R}^M$, which can be directly correlated against physical observables evaluated on the same set of sampled spin configurations.

\begin{figure*}[htbp]
    \centering
\includegraphics[width=1.0\linewidth]{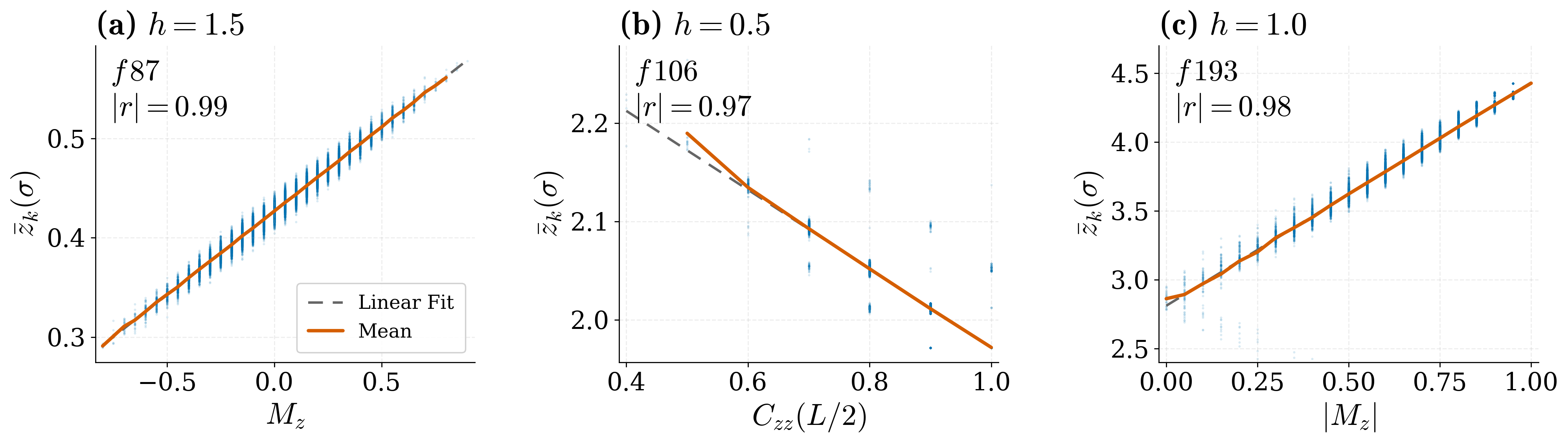}
    \caption{\textbf{Strong Pearson correlation between sparse, autonomously discovered features and physical observables.} \textbf{(a)} Pearson correlation $(|r| = 0.99)$ between sparse feature 87 and the magnetization $M_z$, in the paramagnetic phase ($h=1.5)$; \textbf{(b)} Correlation $(|r|=0.97$) between sparse feature $106$ and half-chain correlator $C_{zz}(L/2)$, in the ferromagnetic phase $(h=0.5)$; and \textbf{(c)} Correlation $(|r|=0.98)$ between the absolute magnetization $|M_z|$ and feature 193, at the critical point ($h=1.0)$. Across all three regimes, despite being trained without any physical labels, the SAE successfully discovers features from residual-stream activations that are strongly aligned with physically meaningful observables.}
    \label{fig:correlation}
\end{figure*}

For the TFIM, we consider three physically motivated observables. Naturally, we consider the order parameter of the model, namely the magnetization
\begin{align}
  M_z(\boldsymbol\sigma) &= \frac{1}{L} \sum_{i=1}^L \sigma_i^z,
\end{align}
together with the absolute magnetization $|M_z|$.

We also probe long range order using the translationally averaged half-chain correlator:
\begin{equation}
    C_{zz}(L/2) = \frac{1}{L} \sum_i \sigma_i^z \sigma_{i+L/2}^z,
\end{equation}
where site indices are modulo $L$ due to periodic boundary conditions. $C_{zz}$ probes spin-spin order and saturates in the ferromagnetically ordered phase.

For each observable $O \in \{M_z,\,|M_z|,\, C_{zz}(L/2)\}$, we evaluate $O^{(m)} \equiv O(\boldsymbol \sigma^{(m)})$ for all sampled spin configurations. We collect the results, forming vectors $\mathbf{O} \in \mathbb{R}^M$. We then compute the correlation between each mean-pooled SAE feature $\mathbf{Z}_k$ and physical observables $\mathbf{O}$:
\begin{equation}
  r_k{(O)} = \mathrm{corr}(\mathbf{Z}_k,\, \mathbf{O}).
\end{equation}

We use the Pearson correlation coefficient, which is defined as:
\begin{equation}
    r_k(O) := \frac{\sum_m \bigl(Z_k^{(m)}-\overline{Z_k}\bigr)\bigl(O^{(m)}-\overline{O}\bigr)}
         {\sqrt{\sum_m \bigl(Z_k^{(m)}-\overline{Z_k}\bigr)^2}\,\sqrt{\sum_m \bigl(O^{(m)}-\overline{O}\bigr)^2}},
\end{equation}
where the overline denotes an average over the $M$ samples. A value $|r_k(O)| = 1$ indicates perfect linear correlation between the $k$th feature and the observable $\mathbf{O}$, while $r_k(O)=0$ indicates no linear correlation. For each observable $O$, we compute $r_k(O)$ for all $d_\text{SAE}$ features and rank the features by $|r_k(O)|$. We denote the maximally correlated feature by $k^*$. We emphasize that we do not claim a one-to-one correspondence between SAE features and physical observables. In this work, we focus on the maximally correlated feature direction $k^*$ because our primary goal is to establish that the residual stream contains at least one interpretable direction corresponding to physical observables.

As shown in Fig.~\ref{fig:correlation}, the SAE autonomously discovers sparse features that correlate strongly with physical observables across the TFIM phase diagram. In the paramagnetic phase ($h=1.5)$, feature 87 correlates with the magnetization $M_z$ at $r=0.99$. In the ferromagnetic phase ($h=0.5$), the autonomously discovered feature 106 correlates with the half-chain correlator $C_{zz}$ at $r=-0.97$, despite $C_{zz}$ being a non-linear observable in the spin configuration. The negative sign of the correlation indicates that the learned feature direction is aligned with $-C_{zz}$ instead of $C_{zz}$. Remarkably, even at the critical point $h=1.0$, the SAE method successfully discovers a feature ($f193$) that strongly correlates with $|M_z|$. 

These remarkably high correlations suggest that the SAE features act as
nearly linear internal coordinates for the corresponding physical observables. Importantly, neither the NQS nor the SAE is trained using these observables as labels: the NQS is trained only through variational energy minimization, while the SAE is trained only to reconstruct residual-stream activations in a sparse way. To confirm that the observed correlations are not artifacts of the transformer or the SAE architectures, we repeat the same analysis for a randomly initialized NQS in Appendix~\ref{app:random_baseline}. In that case, no feature shows any meaningful correlation with physical observables. Thus, the feature-observable alignment emerges from the trained NQS representation rather than from architectural bias.

\subsection{Observable Steering through Sparse Features \label{sec:steering}}
In the previous section, we demonstrated that SAEs identify sparse features that correlate strongly with physical observables across different phases of the TFIM. Although these correlations are highly suggestive, they do not by themselves establish that the features actually \emph{control} the observables predicted by the NQS. To probe causation directly, we perform \textit{targeted intervention}~\cite{causal_intervention} on the NQS: we rescale the top-correlating ($k^*$th) feature and measure how the corresponding observable responds. Related interventions have been successful in natural language processing, steering the behaviors of large language models such as GPT models and Claude 3 Sonnet~\cite{claude_interpretable, GPT_steering}. Similar ideas have also been applied in scientific domains to amplify interpretable concepts in protein and weather models~\cite{ProteinLM_SAE1, weather_model}. 

In our setting, we carry out intervention through \textit{activation steering}~\cite{activation_engineering, activation_steering}.
For each configuration $\boldsymbol\sigma^{(m)}$, we modify the corresponding final-layer residual stream activations
$\mathbf{h}^{(m)}$ in three steps. First, the activation $\mathbf{h}^{(m)}$ is encoded to the sparse activation strengths $\mathbf{z}^{(m)}$ by the SAE encoder (Eq.~\ref{eq:SAE_encoding}). Next, the activation strength of the top-correlated feature is rescaled, $z_{k^*} \rightarrow \widetilde{z_{k^*}} \equiv \alpha z_{k^*}$, while all other activations $k \neq k^*$ remain unchanged. Finally, the modified sparse activation vector $\widetilde{\mathbf{z}}$ is passed through the SAE decoder (Eq.~\ref{eq:SAE_decoding}), to reconstruct a modified residual stream $\widetilde{\mathbf{h}}$, which is then passed through the unembedding layer to define a modified NQS $\widetilde{{\psi}}$. The scale factor $\alpha$ therefore controls the strength of this intervention: $\alpha = 1$ leaves all features unchanged; while $\alpha < 1$ and $\alpha > 1$ suppress and amplify the selected feature $k^*$, respectively. The concept of feature-steering procedure is illustrated in Fig.~\ref{fig:illustration}(c).

\begin{figure}
    \centering
    \includegraphics[width=0.9\linewidth]{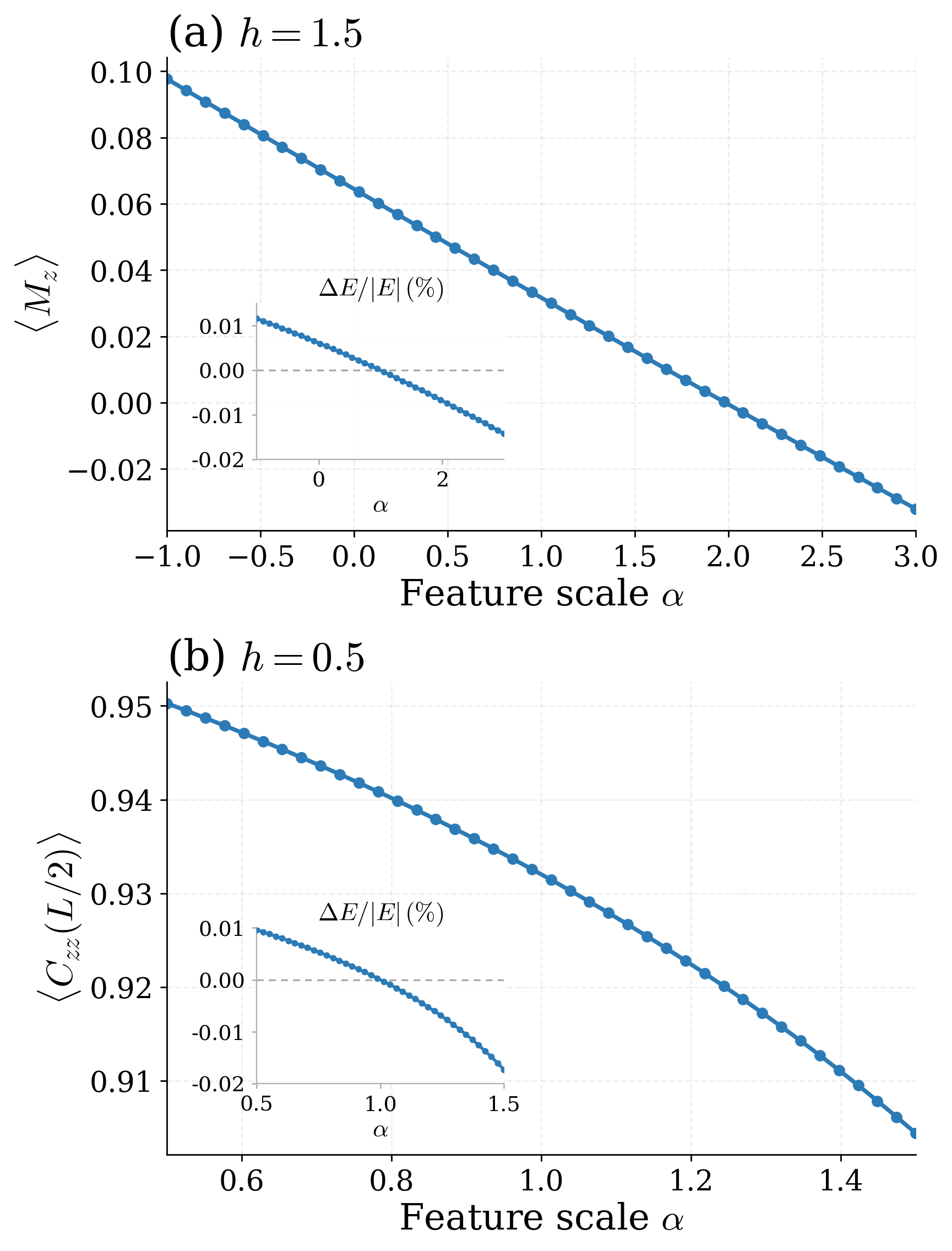}
    \caption{\textbf{Causal steering of physical observables through sparse-feature interventions.} Rescaling the SAE feature with the strongest correlation to a given physical observable by a factor $\alpha$ smoothly and monotonically tunes the correlating observable, while leaving the variational energy almost unchanged. \textbf{(a)} Magnetization $\left< M_z \right>$ as a function of feature scale $\alpha$ in the paramagnetic phase ($h=1.5$). \textbf{(b)} Half-chain correlator decays monotonically with $\alpha$ in the ferromagnetic regime ($h=0.5$). \textit{Insets}: relative energy deviation ($\%$) over the same range of $\alpha$, remaining below $0.02\%$ in both cases. The minimal change of the variational energy indicates that feature steering modifies the target observable in a controlled way, without significantly perturbing the variational state.}
    \label{fig:steering}
\end{figure}

Fig.~\ref{fig:steering} shows how the magnetization $M_z$ and the half-chain correlator $C_{zz}(L/2)$ respond to the feature scale $\alpha$ in the paramagnetic $(h=1.5)$ and ferromagnetic $(h=0.5)$ phases, respectively. In both cases, rescaling the top-correlated feature produces a smooth and monotonic shift in the corresponding physical observables. At the same time, the variational energy remains almost unchanged, with a relative energy deviation $|\Delta E/E|$ remaining below $0.02 \%$ in both cases. This indicates that the intervention is localized in representation space: modifying a single sparse feature can steer the target observable alone. The identified features $k^*$ therefore do not merely correlate with physical observables, but play a causal role in how these observables are represented by the NQS.

There is a crucial distinction between previous intervention studies and our case. For large language models and protein language models, evaluating the effect of steering requires choosing a behavioral metric~\cite{mech_interp_review1, mech_interp_review2, mech_interp_review3}, which can by itself be subjective. In contrast, interventions on sparse features in NQS can be objectively measured by expectation values of unambiguously defined physical observables, with no interpretive ambiguity.

We note that there is an important subtlety about sampling. To isolate the effect of feature steering itself, we evaluate the steering experiment using the \textit{same} set of sampled spin configurations $\{ \boldsymbol \sigma^{(m)}\}$ for all values of $\alpha$. All configurations are thus drawn from the original Born distribution $|\psi|^2$. As a result, $\alpha$ cannot be taken arbitrarily large or small; otherwise the modified Born distribution $|\widetilde{\psi}|^2$ would differ significantly from $|\psi|^2$, and the fixed sample set will no longer provide reliable estimates of expectation values. In Appendix~\ref{app:importance_sampling}, we quantify the range of admissible $\alpha$ and show that the values of $\alpha$ considered in Fig.~\ref{fig:steering} are well within the regime where importance-sampling estimates are reliable.

Such feature-based steering also has potential practical applications. For example, in situations where the exact ground state is unknown, but certain observables are available experimentally, the sparse features identified by SAE provide an interpretable handle for adjusting the NQS toward desired physical behavior, without changing diagnostics such as the variational energy. Importantly, this control acts directly on the residual stream: once a sparse feature strongly associated with a target observable has been identified, the observable can be tuned by simply adjusting a single scalar coefficient immediately before the unembedding layer. The post-training, low-dimensional intervention differs from other approaches such as fine-tuning~\cite{model_param1, transformer_nqs_zhang, fine_tuning_nqs, noqs, unp}, which would require additional optimization and backward passes and can be substantially more expensive for larger models. 

\begin{comment}
The TFIM Hamiltonian commutes with the global spin-flip operator $\mathcal{F}:\sigma_i\to-\sigma_i$,
so the exact ground state satisfies $|\psi(\boldsymbol\sigma)| = |\psi(-\boldsymbol\sigma)|$
(up to spontaneous symmetry breaking below $h_c$).
For each feature $k$, the site-averaged activations on the original and
flipped configurations are decomposed into symmetric and antisymmetric parts:
\begin{align}
  S_k(\boldsymbol\sigma) &= \tfrac{1}{2}\bigl[Z_k(\boldsymbol\sigma) + Z_k(-\boldsymbol\sigma)\bigr],\\
  A_k(\boldsymbol\sigma) &= \tfrac{1}{2}\bigl[Z_k(\boldsymbol\sigma) - Z_k(-\boldsymbol\sigma)\bigr].
\end{align}
The antisymmetry score
\begin{equation}
  \eta_k = \frac{\mathbb{E}|A_k|}{\mathbb{E}|S_k| + \mathbb{E}|A_k| + \varepsilon} \in [0,1]
\end{equation}
measures the fraction of a feature's activation that flips sign under $\boldsymbol\sigma\to-\boldsymbol\sigma$.
Features with $\eta_k \approx 1$ are purely antisymmetric (order-parameter-like);
those with $\eta_k \approx 0$ are purely symmetric (e.g.\ domain-wall counting).
A complementary linear antisymmetry measure is given by
$\mathrm{corr}(Z_k(\boldsymbol\sigma),\,-Z_k(-\boldsymbol\sigma))$.
\end{comment}

\subsection{Real-Time Dynamics}
In this section, we extend our analysis to transformer NQS that capture real-time dynamics. Compared to the ground state analysis, this setting is more challenging, as both the quantum state and associated Born distribution evolve continuously in time. It is not obvious whether a fixed set of sparse features can track physical observables throughout the entire trajectory.

To test this, we study real-time dynamics under the TFIM. 
The system is initialized in the ferromagnetically aligned state
\begin{equation}
    |\psi_0\rangle = |\!\downarrow\downarrow\cdots\downarrow\rangle,
\end{equation}
and evolved under the TFIM Hamiltonian in the paramagnetic regime with $h = 1.5$. As discussed in Sec.~\ref{sec:NQS}, the time-dependent state $\ket{\psi(t)}$ is represented by a transformer NQS with time-dependent parameters $\theta(t)$, optimized using the TDVP over the interval $t \in [0, T]$ 
with $TJ = 0.5$.

The SAE analysis proceeds analogously to the ground state case. At each time step $t$, 
we draw $M$ spin configurations $\boldsymbol \sigma^{(m)}(t)$ from the instantaneous Born distribution 
$|\psi(\boldsymbol \sigma; t)|^2$ and collect the final-layer residual stream 
activations $h^{(m)}(t)$. We then pool together the sampled spin configurations and activation vectors across all times, and train a \emph{single} SAE on the full dynamical dataset. Crucially, the SAE dictionary is \textit{not} trained separately at each time step. Instead, the learned feature directions $\mathbf{f}_k$ remain fixed, while the activation strengths $\mathbf{z}_k(t)$ vary over time.

For each time step, we then analyze the correlation between physical observable $\mathbf{O}(t)$ and mean-pooled activations $\mathbf{Z}_k(t)$ for all features, just as in the ground state setting (Sec.~\ref{sec:gs_correlation}). This construction asks whether a sparse feature continues to align with the same observable
during the trajectory. Features that maintain a high correlation $|r(t)|$ over time 
can therefore be interpreted as the internal coordinates that the NQS uses to represent the dynamics of that given observable.

\begin{figure}
    \centering
\includegraphics[width=0.9\linewidth]{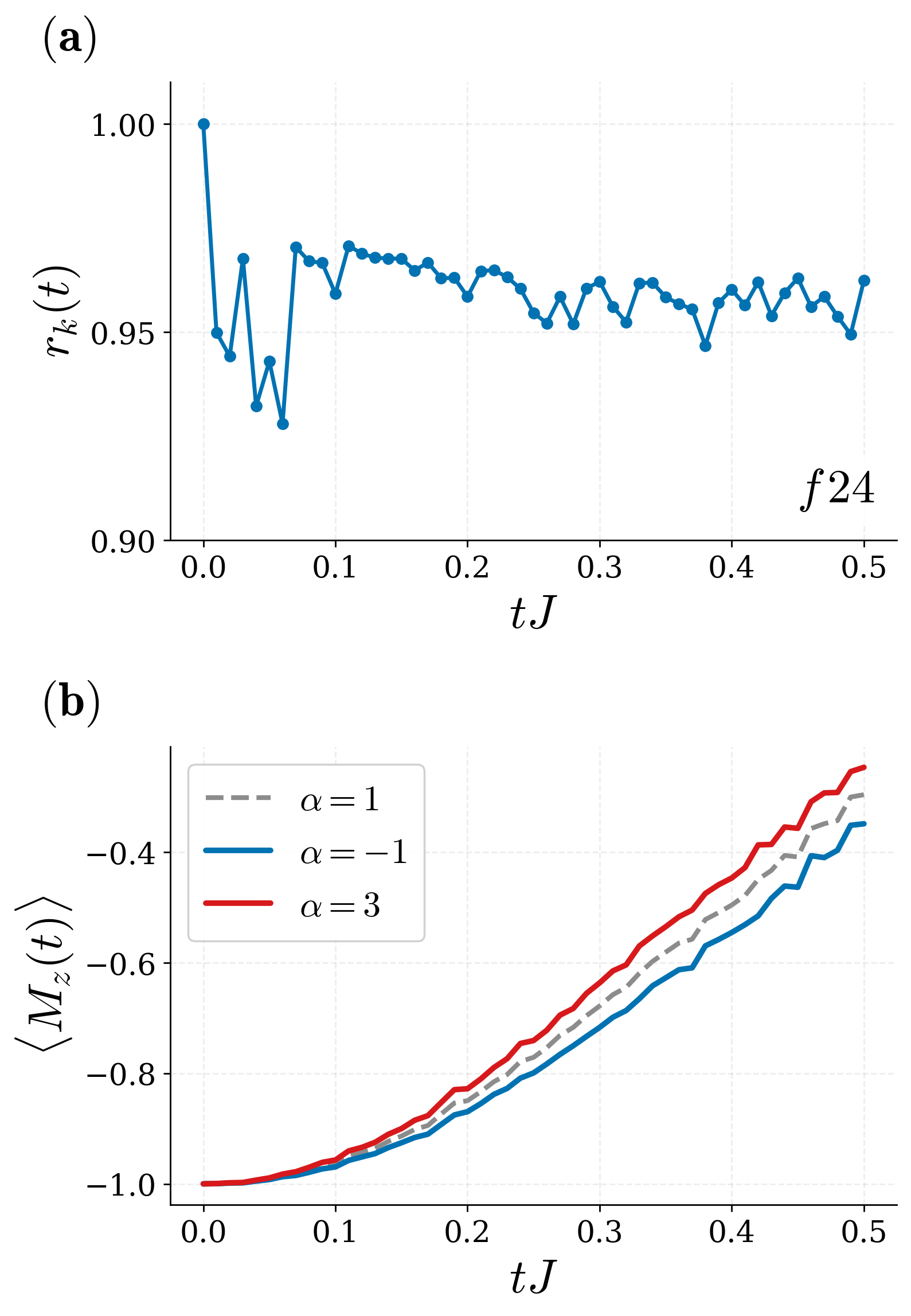}
    \caption{\textbf{Feature identification and steering in real-time dynamics. (a)} Pearson correlation between the activation strength $z_{k}(t)$ of sparse feature $f24$ and the instantaneous magnetization $M_z(t)$, during time-evolution under the TFIM. The correlation remains high $(>0.92)$ throughout the full interval, indicating that a single, static feature direction tracks magnetization dynamics over the entire trajectory. \textbf{(b)} Dynamical steering of the magnetization $\langle M_z(t)\rangle$ through  rescaling the activation strength, $\mathbf{z}_{k^*}(t) \rightarrow \alpha \mathbf{z}_{k^*}(t)$.  Compared to the unmodified case ($\alpha=1$), suppressing ($\alpha=-1$) and amplifying ($\alpha=3$) the single feature produce a clear and systematic effect on the magnetization dynamics.}
    \label{fig:time_dependent}
\end{figure}

As shown in Fig.~\ref{fig:time_dependent}(a), a single sparse feature ($f24$) maintains a strong correlation with the instantaneous magnetization $M_z(t)$ throughout the entire evolution. Specifically, the Pearson correlation $|r(\tilde{z}_{k^*}, 
M_z)|$ remains above 0.92 for all $tJ \in [0, 0.5]$, with an average correlation value close to 0.95. Thus, this feature not only captures the magnetization near the initial polarized state, but also \textit{continues to} track the magnetization faithfully, even as the magnetization evolves significantly away from its initial value. Remarkably, time and magnetization are not explicitly input into the SAE, which must infer from the residual-stream activation vectors alone which direction is associated with the magnetization dynamics.

We next test whether feature-steering is applicable to the dynamical state. We intervene on the identified SAE feature by scaling its activation strength across all times, $\mathbf{z}_{k^*}(t) \rightarrow \alpha \,\mathbf{z}_{k^*}(t)$, while leaving all remaining feature activations unchanged. Fig.~\ref{fig:time_dependent}(b) shows the resulting time-dependent magnetization $\langle M_z(t) \rangle$ for three representative values of the scale factor: $\alpha = 1$ (original state), $\alpha = -1$ (feature suppressed), and $\alpha = 3$ (feature amplified). Rescaling the single feature produces a clear and systematic effect on the magnetization: amplifying $f24$ accelerates the departure from the initial magnetization, while suppressing $f24$ keeps the state more polarized. The separation between the resulting trajectories is persistent throughout the full time interval, indicating that the feature captures magnetization dynamics over the entire evolution instead of at isolated times.

These results demonstrate that the SAE analysis is not limited to static wavefunction representations. Even when the NQS parameters evolve in time, the SAE is able to learn a static basis, whose time-dependent activation strengths both correlate with and causally steer physical observables over the time interval. This suggests that the transformer NQS encodes dynamical physical information in terms of a \textit{static} internal coordinate system. More generally, the constant scaling considered here is the simplest intervention protocol. One could, for example, introduce a time-dependent control function $\alpha(t)$, and implement a more refined steering of the relevant direction, $\mathbf{z}_{k^*}(t) \rightarrow \alpha(t) \mathbf{z}_{k^*}(t)$. In such cases, the sparse feature would serve as a dynamical control knob for targeted manipulation of time-dependent neural quantum states.

\section{2D Model: Heisenberg Antiferromagnet\label{sec:heisenberg_afm}}
While the TFIM provides a controlled setting for benchmarking our approach, it is also important to test whether the SAE analysis extends beyond one-dimensional, integrable models. In this section, we turn to a more challenging quantum many-body system, the spin-$1/2$ Heisenberg
Antiferromagnet (AFM) model in two dimensions~\cite{heisenberg_afm, heisenberg_afm_review}. This model is a paradigmatic model for quantum magnetism. Its ground state exhibits antiferromagnetic long-range correlations, as well as strong quantum fluctuations, making it a significantly more challenging benchmark for testing whether SAE features can identify physically meaningful observables.

Concretely, we consider the Heisenberg AFM on a square lattice of size $L_x \times L_y$ with
periodic boundary conditions. The Hamiltonian is:
\begin{equation}
    H = J \sum_{\langle ij \rangle} \left(
          \sigma_i^x \sigma_j^x + \sigma_i^y \sigma_j^y + \sigma_i^z \sigma_j^z
        \right),
\end{equation}
where $\langle ij \rangle$ denotes nearest-neighbor bonds on the square lattice, $\sigma_i^x, \sigma_i^y, \sigma_i^z$ are Pauli
operators on site $i$. We set $J > 0$, corresponding to antiferromagnetic exchange. 

The order parameter is the \emph{staggered} magnetization, which is defined as:
\begin{equation}
    M_\mathrm{stag} = \frac{1}{L} \sum_i (-1)^{x_i + y_i} \sigma_i^z,
\end{equation}
where $i = (x_i, y_i)$ labels the site coordinate on the square lattice, $L = L_x
L_y$ is the total number of sites, and the factor $(-1)^{x_i+y_i}$ assigns opposite signs to the two sublattices. 
%For a classical antiferromagnet, spins on the $A$-sublattice ($x_i + y_i = 0 \mod 2$) point up and spins on the $B$-sublattice ($x_i + y_i = 1\mod 2$) point down, giving $|M_\mathrm{stag}| = 1$. In quantum systems, however, the ground state of the 2D Heisenberg AFM has $M_\mathrm{stag} \approx 0.307$~\cite{sandvik1997} due to quantum fluctuation. The staggered magnetization is one of our checks on when NQS has converged to the ground state, as well as the observable we correlate against sparse features.

As in the TFIM case, we train a transformer NQS to approximate the ground state, through minimizing the variational energy, for a lattice of size $6 \times 6$. We determine training convergence by benchmarking the variational energy against the ground state energy density determined from Quantum Monte Carlo (QMC)~\cite{sandvik1997}. The SAE analysis proceeds in the same way as Sec.~\ref{sec:gs_correlation}: 
after training, we sample $M$ spin configurations and collect the corresponding final-layer residual-stream activation vectors; we then train an SAE to extract sparse features from the activations; and finally compute the Pearson correlation between $\mathbf{Z}_k$ and $\mathbf{M}_\text{stag}$, ranking the features by the correlation strength.

\begin{figure}
    \centering
    \includegraphics[width=0.9\linewidth]{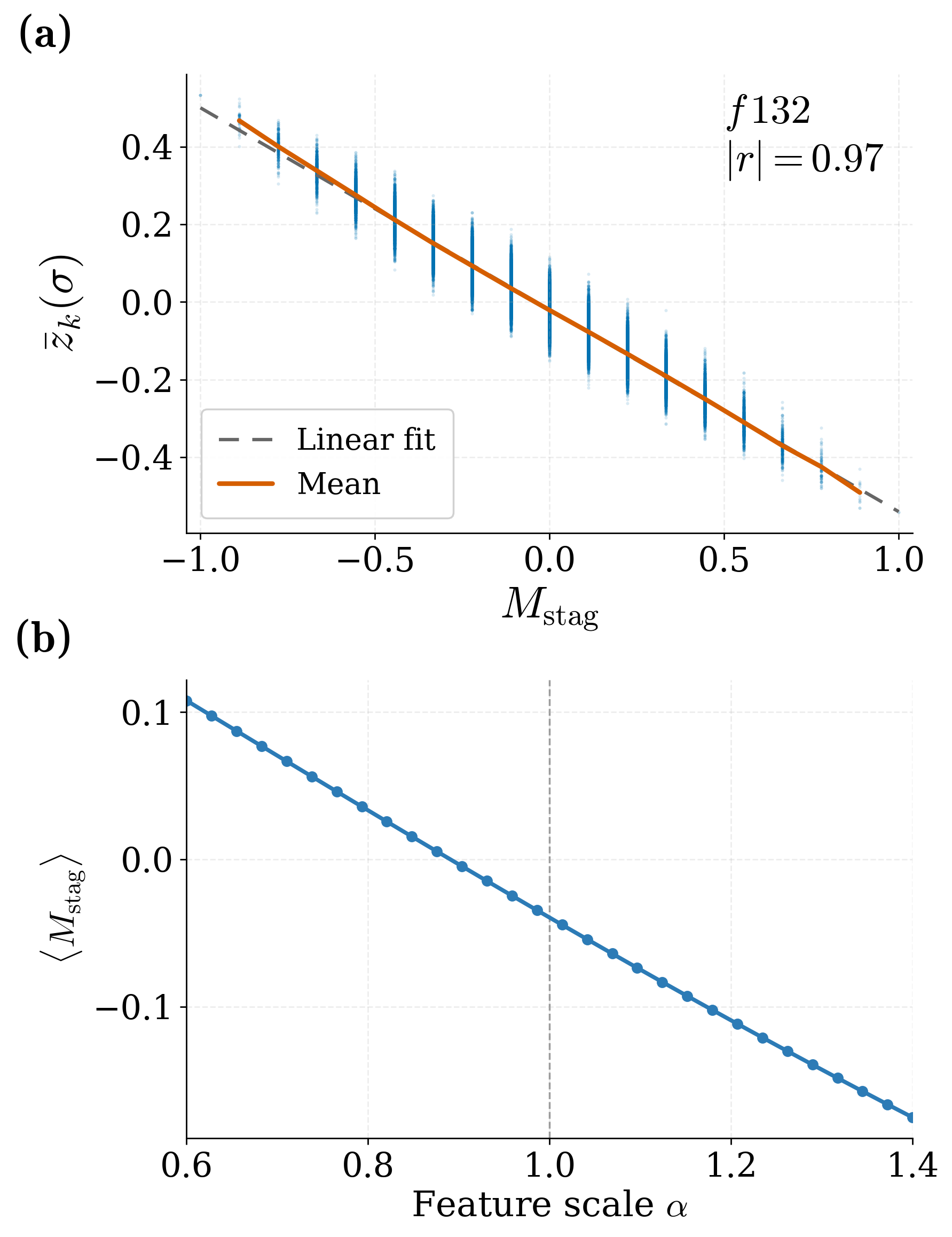}
    \caption{\textbf{SAE identifies a feature that strongly correlates with and causally controls staggered magnetization in the two-dimensional Heisenberg antiferromagnet. (a)} Strong correlation between SAE feature activation and staggered magnetization $M_\text{stag}$, with $r=-0.97$. \textbf{(b)} Feature steering of the staggered magnetization. Rescaling the activation of $f132$ by a factor $\alpha$ produces a smooth and monotonic change in $\langle M_\text{stag} \rangle$, showing that the feature correlating with $M_\text{stag}$ also controls it.}
    \label{fig:AFM}
\end{figure}

Fig.~\ref{fig:AFM}(a) shows that the SAE identifies a sparse feature ($f132$) that is strongly correlated with the staggered magnetization, with a correlation coefficient $|r|=0.97$ (the negative slope means that the feature is aligned with $-M_\text{stag}$ instead of $+M_\text{stag}$). To probe whether the activation strengths causally control $M_\text{stag}$, we follow the same activation-steering procedure used in Sec.~\ref{sec:steering}: rescaling the activation strength of $f132$ by $\alpha$, while leaving all other feature strengths unchanged, and modifying the residual stream of transformer NQS. As shown in Fig.~\ref{fig:AFM}(b), increasing and decreasing the strength of the top-correlating feature monotonically changes the predicted staggered magnetization, establishing that the SAE again successfully identifies a sparse feature that tracks the staggered order parameter.

Our results establish that the interpretability approach holds beyond simple, one-dimensional systems, and also beyond simple quantities in terms of spin configurations. Indeed, compared to magnetization in the TFIM, the staggered magnetization is not a uniform sum over all sites, but a spatially structured quantity that distinguishes the two sublattices. The fact that a single SAE feature tracks this quantity indicates that the transformer NQS can internally form representations for quantities beyond global spin information. Even in a two-dimensional interacting quantum magnet, the trained NQS model organizes physically relevant information in sparse residual stream directions, by encoding the sublattice pattern associated with the antiferromagnetic long-range order into its activation patterns.

\section{Discussion\label{sec:discussions}}
\textit{Do neural networks dream of Schr\"odinger's Cat?} In this work, we show that for neural quantum states, the answer is affirmative. Using sparse autoencoders to analyze the residual-stream of transformer NQS, we have successfully identified physically interpretable features from the NQS's internal representation. The autonomously identified features not only strongly correlate with physical observables, but in fact \textit{causally} control them. By adjusting the strengths of the top-correlating feature activations, we can tune the corresponding observable in a controlled way. Importantly, our analysis holds for both real-time dynamics, where a \textit{static} feature tracks time-dependent magnetization across the entire time window, and for a two-dimensional, non-integrable quantum antiferromagnet.

Our results provide evidence that NQS models, trained solely on variational objectives, form internal representations that encode relevant physical quantities of the system. Furthermore, our approach provides a diagnostic tool for NQS. If order parameters and relevant observables are represented by internal coordinates, this provides evidence that the NQS could have learned a physically sensible representation; conversely, if such correlations are absent, it could help diagnose failure modes of the model and potentially point to better architectures or physical constraints. 

Our findings open up a frontier of future directions. In this work, we have focused on the transformer architecture exclusively. Whether other neural architectures, such as Restricted Boltzmann Machines (RBM)~\cite{carleo2017} or recurrent neural networks~\cite{autoregressive_nqs2}, organize physical information in a similar way remains open. Such comparisons could help identify architectural choices that make NQS more interpretable or physically faithful. 

The SAE model in this work also only decomposes activation vectors from the final-layer residual stream. This is a natural starting point, as these activations feed directly into the unembedding layer; however, a systematic, layer-resolved analysis of the NQS and especially the self-attention mechanism could also carry physical information. For instance, do attention weights distinguish short-range from long-range order across phase transitions? Probes of this kind, using interpretability tools, could potentially provide novel diagnostics of phase boundaries.

As SAE was originally designed for foundation models, it would be interesting to see if our analysis could provide further insight on foundational NQS models, which are trained across families of Hamiltonians or driving protocols~\cite{noqs, unp, foundational_nqs, liangfu_foundation}. For example, certain sparse features could represent order parameters across phase transitions. If such features exist, they could provide insights into how foundation NQS models organize information across families of Hamiltonians rather than individual wavefunctions.

More broadly, our work suggests a two-way exchange between the communities of neural quantum states and mechanistic interpretability. NQS provide a setting where interpretability claims can be tested against clearly defined physical observables and exact results; conversely, mechanistic interpretability provides tools for opening the black box of variational neural wavefunctions. As NQS take on a larger role in quantum simulations, understanding their internal representations will become essential for diagnosing failures, building trust, and potentially even extracting novel physical insights from trained models.

\section*{Acknowledgments}
We thank Junkai Dong, Toni Liu, Yang Peng, and Miles Stoudenmire for helpful discussions, and especially Yang Peng and Junkai Dong for reading through the manuscript. The authors acknowledge the use of ChatGPT 5.5 and Sonnet 4.6 for code writing and manuscript polishing. ZQ gratefully acknowledges support from the Simons Center for Geometry and Physics, Stony Brook University at which some of the research for this work was performed during the program \emph{Complexity, Information, and Tractable Simulations of Quantum Many-body Dynamics}.

\bibliography{main}

\appendix
\section{Details of Transformer Architecture \label{app:transformer_architecture}}
The transformer architecture~\cite{attention_is_all_I_need} has recently become a powerful ansätz for quantum wavefunctions~\cite{transformer_frustrated, transformer_nqs_zhang, foundational_nqs, liangfu_foundation}. Its self-attention mechanism allows the wavefunction representation to capture long-range correlations in many quantum states. In this work, we use a decoder-only autoregressive transformer as the NQS architecture.

As illustrated in Fig.~\ref{fig:illustration}(a), inputs to the transformer are spin configurations $\boldsymbol \sigma = (\sigma_1, \sigma_2, \dots, \sigma_L)$; these are the transformer's tokens. The spin tokens are first lifted to a latent space with dimension $d_\text{model}$, via an embedding operation:
\begin{equation}
    \mathbf{e}_i = W_E \, \sigma_i,
\end{equation}
where $W_E$ is an embedding weight matrix shared across all sites. Each site is also augmented with a learnable positional encoding $\mathbf{p}_i \in \mathbb{R}^{d_\text{model}}$. The input to the transformer blocks is therefore:
\begin{equation}
    \mathbf{x}_i = \mathbf{e}_i +\mathbf{p}_i, \quad i\in \{1, 2, \dots, L\}.
\end{equation}

The sequence $\{\mathbf{x}_i\}_{i=1}^L$ is then passed through $L_T$ transformer layers. Each layer consists of two parts: a multi-head self-attention layer followed by a feedforward multi-layer perceptron, with residual connections around both. For notational simplicity, we suppress the transformer-block index in the following.

Let $X \in \mathbb{R}^{L \times d_\text{model}}$ denote the residual stream entering a transformer block. The first sublayer is masked multi-head self-attention. For each attention head $a=1,\ldots,n_h$ (where $n_h$ is the number of heads), the normalized residual stream is projected into queries, keys, and values:
\begin{equation}
    Q^{(a)} = X W_Q^{(a)}, 
    \quad 
    K^{(a)} = X W_K^{(a)}, 
    \quad 
    V^{(a)} = X W_V^{(a)} ,
    \label{eq:self_attn_proj}
\end{equation}
where
\begin{equation}
    W_Q^{(a)}, W_K^{(a)}, W_V^{(a)}
    \in
    \mathbb{R}^{d_\text{model} \times d_h}
\end{equation}
are the projection weight matrices, and $ d_h = d_\text{model}/n_h$ is the dimension per attention head. At each position, the query determines what information that position is seeking; the key determines whether this position is relevant to queries from all other positions; and the value contains the information that can be passed to other positions.

For each head, we compute the attention score
\begin{equation}
    S^{(a)}= \frac{
        Q^{(a)} {K^{(a)}}^T}{
        \sqrt{d_h}}+ \mathcal{M},
    \label{eq:attention_scores}
\end{equation}
where $\mathcal{M}\in\mathbb{R}^{L\times L}$ is the mask:
\begin{equation}
    \mathcal{M}_{ij}
    =
    \begin{cases}
        0, & j \leq i, \\
        -\infty, & j > i.
    \end{cases}
    \label{eq:causal_mask}
\end{equation}
The mask is \textit{causal} in the sense that it sets the weight of every future position to zero, so the representation at position $i$ can only use information from previous spin positions $j\leq i$.

The attention matrix is obtained by applying a row-wise softmax,
\begin{equation}
    A^{(a)}_{ij} = \frac{
        \exp \left(S^{(a)}_{ij}\right)}{
        \sum_{m=1}^{L} \exp \left(S^{(a)}_{im}\right)}.
    \label{eq:attention_weights}
\end{equation}

The output of head $a$ is a weighted sum of value vectors,
\begin{equation}
    O^{(a)}
    =
    A^{(a)} V^{(a)}
    \in
    \mathbb{R}^{L \times d_h}.
    \label{eq:attention_output_single_head}
\end{equation}
The outputs of all heads are concatenated and projected back to the embedding dimension:
\begin{equation}
    \mathrm{Attn}(X)
    =
    \mathrm{Concat}
    \left(
        O^{(1)},\ldots,O^{(n_h)}
    \right)
    W_O
    +
    \mathbf{b}_O,
    \label{eq:self_attn_concat}
\end{equation}
where $W_O\in\mathbb{R}^{d_\text{model}\times d_\text{model}}$ and $\mathbf{b}_O\in\mathbb{R}^{d_\text{model}}$. The self-attention result is added back to the input, $X' = X + \text{Attn}(X)$.

The second sublayer is a position-wise feedforward network that mixes the internal feature channels. For the input matrix $X'\in\mathbb{R}^{L\times d_\text{model}}$, the feedforward network is
\begin{equation}
    \mathrm{FFN}(X')
    = \sigma
    \left(
        X' W_1 + \mathbf{b}_1
    \right)
    W_2
    +
    \mathbf{b}_2,
    \label{eq:ffn}
\end{equation}
where $W_1\in\mathbb{R}^{d_\text{model}\times d_f}, \,W_2\in\mathbb{R}^{d_f\times d_\text{model}}$ are the weight matrices of each layer. Here $d_f$ is the feedforward hidden dimension. The feedforward network first expands the channel dimension from $d_\text{model}$ to $d_f$, applies the nonlinear function $\sigma$, and then projects back to $d_\text{model}$.

The output of the transformer layer is therefore
\begin{equation}
    X_{\mathrm{out}}
    =
    X'
    +
    \mathrm{FFN}
    \left(
        X'
    \right).
    \label{eq:ffn_residual_update}
\end{equation}
The same structure is repeated across all $L_T$ transformer layers. After the final layer, we obtain the final-layer residual stream
\begin{equation}
    H \equiv X^{(L_T)}_\text{out} \in \mathbb{R}^{L \times d_\text{model}}.
\end{equation}

This is the representation analyzed by the SAE in this work, right before the unembedding layer.
The activation vector at each position, $\mathbf{h}_i \in \mathbb{R}^{d_\text{model}}$, is linearly mapped to logits over the spin dictionary:
\begin{equation}
    \mathbf{l}_i = \mathbf{h}_i W_U + \mathbf{b}_U, \qquad \mathbf{l_i} \in \mathbb{R}^2
\end{equation}
where $\mathbf{l}_i$ is the logit on site $i$, $W_U \in \mathbb{R}^{d_\text{model}\times 2}$ and $\mathbf{b}_U \in \mathbb{R}^2$ are unembedding weight and bias. 

The conditional probability for the physical spin $\sigma_i$ is obtained by applying a softmax to these logits
\begin{equation} p_\theta(\sigma_i=s\mid \sigma_{<i}) = \frac{ \exp(l_{i,s}) }{ \sum_{s'\in \{\pm 1\}}\exp(l_{i,s'}) }, \quad s\in \{-1, +1\}. \label{eq:softmax_conditional} \end{equation} 

A second linear head maps the \emph{same} residual
stream to the conditional phase. The wavefunction amplitude is assembled from both the amplitude and phase.

Thus, the final residual-stream activation vectors are not themselves wavefunction amplitudes. They are first unembedded into logits, and the logits are then normalized by a softmax to produce the autoregressive conditional distributions. The SAE, therefore, does not directly analyze the probability distribution outputted by the transformer NQS, but an internal representation immediately upstream of the unembedding layer.

\section{Hyperparameters of NQS and SAE \label{app:nqs_sae_training}}

To ensure reproducibility, we list the hyperparameters used in the NQS and SAE in Table~\ref{tab:nqs_hyperparameters} and Table~\ref{tab:sae_hyperparameters}, respectively. The same choice of parameters is used across the TFIM ground state, TFIM real-time evolution, and Heisenberg AFM.

\begin{table}[ht]
\centering

\begin{tabular}{ll}
\hline
\textbf{} & \textbf{Hyperparameter} \\ \hline
\textbf{Architecture} & \\
Number of Transformer Layers ($L_T$) & 2 \\
Embedding Dimension ($d_\text{model}$) & 64 \\
Number of Attention Heads ($n_h$) & 4 \\
Feed-forward Dimension ($d_f$) & $4 \times d_\text{model} = 256$ \\
Activation Function $\sigma$ & ReLU \\
\hline
\textbf{Training} & \\
Optimizer & Adam \\
Learning Rate (LR) & $1 \times 10^{-3}$ \\
Batch size & 256 \\
Training Steps & 300 \\
\hline
\end{tabular}
\caption{Hyperparameters for the Transformer NQS. \label{tab:nqs_hyperparameters}}
\end{table}

\begin{table}[ht]
\centering
\begin{tabular}{ll}
\hline
\textbf{} & \textbf{Hyperparameter} \\ \hline
\textbf{Architecture} & \\
Input Dimension (same as $d_\text{model}$) & 64 \\
Expansion Factor & 4 \\
Hidden State Dimension & 256 \\
Activation Function & ReLU \\
Sample Size $(M)$ & 20000 \\
\hline
\textbf{Training} & \\
Optimizer & Adam \\
Learning Rate (LR) & $1 \times 10^{-3}$ \\
SAE Batch size & 1024 \\
Training Steps & 30000 \\
Sparsity Loss Strength $\lambda$ & 3.0 \\
\hline

\end{tabular}
\caption{Hyperparameters for the Sparse Autoencoder (SAE). \label{tab:sae_hyperparameters}}
\end{table}

In addition to the $L_1$ loss in the main text, one may also train a top-K SAE~\cite{TopKSAE}, in which sparsity is imposed directly by retaining only the $K$ largest feature activations and setting the remaining coefficients to zero. We tested both loss functions for training the SAE and found no significant differences in the feature-observable correlations. The results in this work use the $L_1$ penalized loss function.

\section{Admissible Range of Steering Parameter $\alpha$ \label{app:importance_sampling}}
In Sec.~\ref{sec:results}, we demonstrate the ability to control observables by rescaling the top-correlated features by $\alpha$: $\mathbf{z}_{k^*} \rightarrow \alpha \mathbf{z}_{k^*}$. To isolate the effect of this intervention from any resampling artifact, we used the same set of spin configurations sampled from the original (unmodified) Born distribution $|\psi|^2$. As a result, the modified distribution $|\widetilde{\psi}|^2$ must remain sufficiently close to the original distribution, for the fixed sample set to provide reliable estimates.

Observables under the modified wavefunction $\widetilde{\psi}(\alpha)$ are estimated by importance sampling. For each spin configuration, we define the weight
\begin{equation}
  w(\boldsymbol\sigma) = \frac{|\widetilde{\psi}(\boldsymbol\sigma)|^2}{|\psi(\boldsymbol\sigma)|^2}.
  \end{equation}
The expectation value of an observable $O$ under the modified wavefunction is estimated as:
  \begin{equation}
  \langle O\rangle_{\widetilde{\psi}} \approx \frac{\sum_\sigma w(\sigma)O(\sigma)}{\sum_\sigma w(\sigma)}.
\end{equation}

The quality of this estimate depends on whether the weights are well-behaved: if a small number of samples dominate the average, then the estimator has a high variance, and the estimated observables are no longer reliable. We monitor this effect using the effective sample size (ESS),
\begin{equation}
  \mathrm{ESS} = \frac{\bigl(\sum_{\boldsymbol\sigma} w(\boldsymbol\sigma)\bigr)^2}{\sum_{\boldsymbol\sigma} w(\boldsymbol\sigma)^2},
\end{equation}
expressed as a fraction of total sample size $M$. Intuitively, the ESS estimates the effective number of useful samples from the original distribution that would contribute correctly to estimating expectation values under the modified distribution $|\widetilde{\psi}|^2$.
Therefore, when $\mathrm{ESS}/M$ is close to 1, the modified distribution is close to the original one, and nearly all $M$ samples contribute effectively, yielding reliable estimates. As $\alpha$ moves away from 1, the weights become increasingly uneven, and the variance of the estimator increases, leading to incorrect expectation values.

\begin{figure}[!t]
    \centering
    \includegraphics[width=1\linewidth]{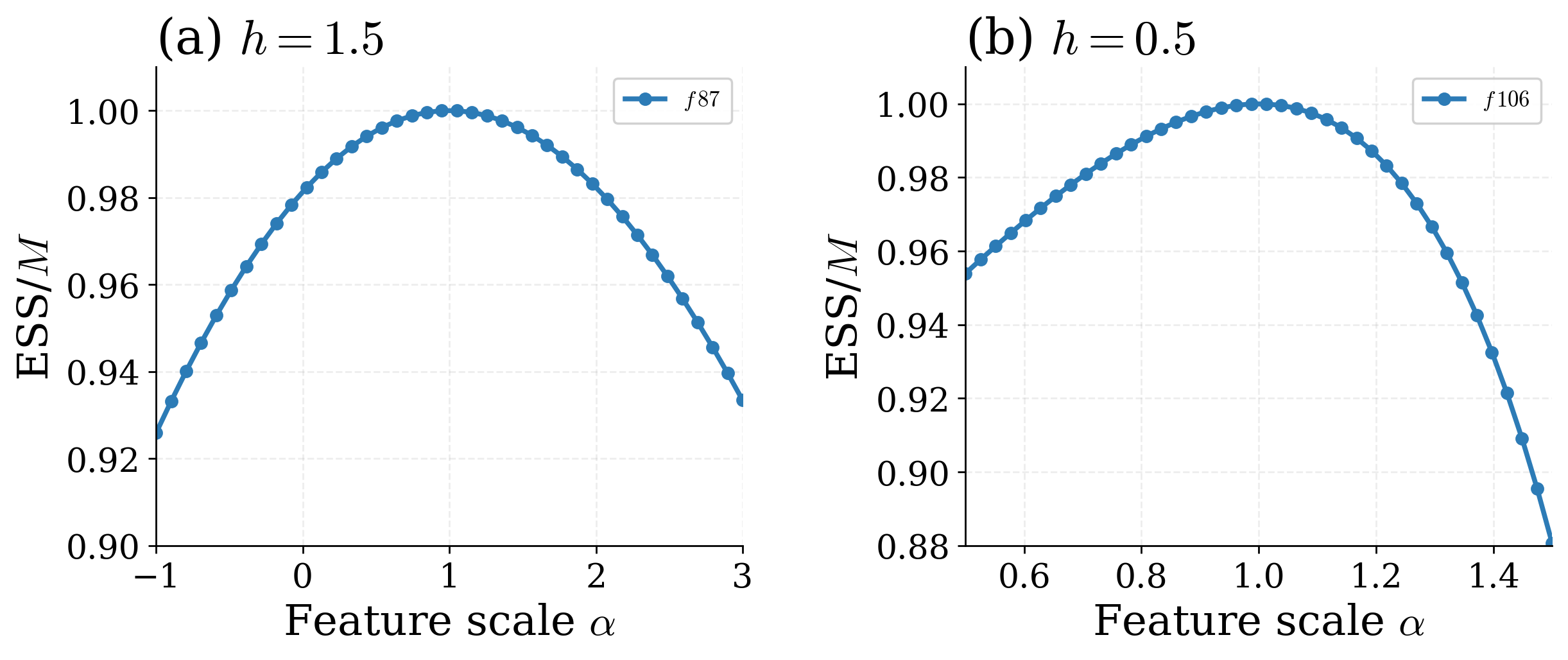}
    \caption{Effective Sample Size (ESS) for feature-steering experiments, expressed as a fraction of total sample size $M$, for the ranges of tuning strength $\alpha$ considered in Fig.~\ref{fig:steering}. The ratio remains above 0.90 in both cases, indicating that the importance sampling estimates are reliable over the steering range considered.}
    \label{fig:ESS}
\end{figure}

Fig.~\ref{fig:ESS} shows the ratio $\mathrm{ESS}/M$ for both the paramagnetic case ($h=1.5)$ and ferromagnetic case ($h=0.5$) for the range of $\alpha$ considered in Fig.~\ref{fig:steering}. For the values of $\alpha$ considered, the ratio ESS$/M$ remains above 0.90, which confirms that the importance-sampling estimates are reliable; therefore, the observed changes in observables are not artifacts of weight degeneracy.

\section{Randomly Initialized NQS \label{app:random_baseline}}
In Fig.~\ref{fig:correlation}, we have demonstrated strong correlation between certain sparse features with physical observables such as magnetization. To verify that this correlation is not an artifact of the transformer architecture, the sampling procedure, or the SAE itself, we repeat the analysis using a \textit{randomly initialized} NQS of the same architecture and hyperparameters as the trained model. The SAE is trained on the final-layer residual-stream activations of this random NQS. We draw the same number of configurations and repeat the same SAE analysis as in the main text.

\begin{figure}
    \centering
    \includegraphics[width=0.8\linewidth]{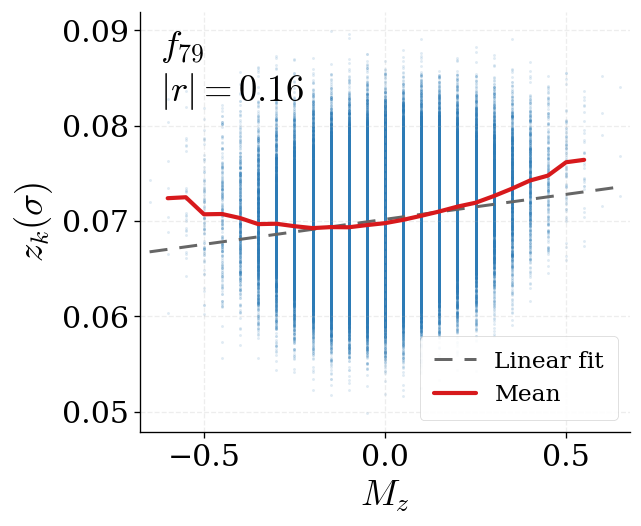}
    \caption{\textbf{Scatter plot of the activations and magnetization for the top-correlated feature of an untrained, randomly initialized NQS model.} The maximum correlation is $|r|=0.16$, in contrast to $|r| = 0.99$ of a model that has converged to the correct ground state (Fig.~\ref{fig:correlation}(a)). This confirms that the strong correlation emerges from NQS training instead of from architectural or SAE bias.}
    \label{fig:random_baseline}
\end{figure}

Fig.~\ref{fig:random_baseline} shows the scatter plot between the \textit{top-correlated feature} ($f79$) and the magnetization $M_z$, in the paramagnetic regime $h=1.5$. The maximum correlation is only $|r|=0.16$, which is in sharp contrast to the trained model, where the top feature tracks magnetization with $|r|=0.99$ (Fig.~\ref{fig:correlation}(a)). This confirms that the high feature-observable correlations observed in the trained models emerge during NQS learning. The NQS must learn an accurate representation of the quantum state for the SAE to recover physically meaningful features.

\end{document}